\documentclass[
]{ceurart}

\usepackage{graphicx}
\usepackage{float} 
\usepackage{latexsym}    %
\usepackage{color}       %
\usepackage{dcolumn}     
\usepackage{bm}          
\usepackage{amssymb}     
\usepackage{hyperref}    
\expandafter\let\csname equation*\endcsname\relax
\expandafter\let\csname endequation*\endcsname\relax
\usepackage{amsmath}
\usepackage{mathtools}          

\newcommand{\be}{\begin{equation}}
\newcommand{\ee}{\end{equation}}

\def\obs#1{{\color{black}{#1}}}

\def\MB#1{{\color{black}{#1}}}
\def\MBB#1{{\color{black}{#1}}}

\newcommand{\bn}{\boldsymbol{n}}

\newcommand{\bx}{\boldsymbol{x}}

\newcommand{\ve}[1]{\ensuremath{\boldsymbol{#1}}}

\begin{document}

\copyrightyear{2020}
\copyrightclause{Copyright for this paper by its authors.
  Use permitted under Creative Commons License Attribution 4.0
  International (CC BY 4.0).}
\conference{The 7th Italian Workshop on Artificial Intelligence and Robotics (AIRO 2020)}

\title{Optimal control of point-to-point navigation in
turbulent time dependent flows using Reinforcement Learning}

\author[1]{M. Buzzicotti}[%
email=michele.buzzicotti@roma2.infn.it,
]
\author[1]{L. Biferale}[%
email=biferale@roma2.infn.it,
]
\address[1]{Dept. Physiscs and INFN, University of Rome Tor Vergata, Via della Ricerca Scientifica 1, 00133 Rome -Italy}

\author[1,2]{F. Bonaccorso}[%
email=fabio.bonaccorso@roma2.infn.it,
]
\address[2]{Center for Life Nano Science@La Sapienza, Istituto Italiano di Tecnologia, 00161 Roma, Italy}

\author[3]{P. {Clark di Leoni}}[%
email=pato@jhu.edu,
]
\address[3]{Department of Mechanical Engineering, Johns Hopkins University, Baltimore, Maryland 21218, USA.}
\author[4]{K. Gustavsson}[%
email=kristian.gustafsson@physics.gu.se,
]
\address[4]{Dept. of Physics, University of Gothenburg, Gothenburg, 41296, Sweden.}

\begin{abstract}
We present theoretical and numerical results concerning the problem to find  the path that minimizes the time to navigate between two given points in a complex fluid under realistic navigation constraints. We contrast deterministic Optimal Navigation (ON) control with stochastic policies obtained by Reinforcement Learning (RL) algorithms. We show that Actor-Critic RL algorithms are able to find quasi-optimal solutions in the presence of either time-independent or chaotically evolving flow configurations. For our application, ON solutions develop unstable behavior within the typical duration of the navigation process, and are therefore not useful in practice. \obs{We first explore navigation of turbulent flow using a constant propulsion speed.} Based on a discretized phase-space, the propulsion direction is adjusted with the aim to minimize the time spent to reach the target. \MBB{Further, we explore a case where additional control is obtained by allowing the engine to power off. Exploiting advection of the underlying flow, allows the target to be reached with less energy consumption. In this case, we optimize a linear combination between the total navigation time and the total time the engine is switched off.} Our approach can be generalized to other setups, for example, navigation under imperfect environmental forecast or with different models for the moving vessel. 
\end{abstract}

\begin{keywords}
  Optimal Control\sep
  Reinforcement Learning \sep
  Turbulence \sep
  Unmanned Navigation
\end{keywords}

\maketitle

\section{Introduction}
Controlling and planning paths of small autonomous marine vehicles
\cite{petres2007path} such as wave and current gliders
\cite{kraus2012wave}, active drifters
\cite{lumpkin2007measuring}, buoyant underwater explorers, and small swimming drones is important for many
geo-physical \cite{lermusiaux2017future} and engineering
\cite{bechinger2016active} applications. In realistic open environments, these vessels are
affected by disturbances like wind, waves and ocean
currents, characterized by unpredictable
(chaotic) trajectories. Furthermore, active control is also limited by engineering and budget aspects as for the important case of unmanned drifters for oceanic exploration 
\cite{centurioni2018drifter,roemmich2009argo}. The problem of (time) optimal point-to-point navigation in a flow, known as Zermelo's problem \cite{zermelo1931}, is interesting {\it per se} in the framework of Optimal Control Theory \cite{bryson2018}.
\begin{figure}
\centering
\includegraphics[scale=0.42]{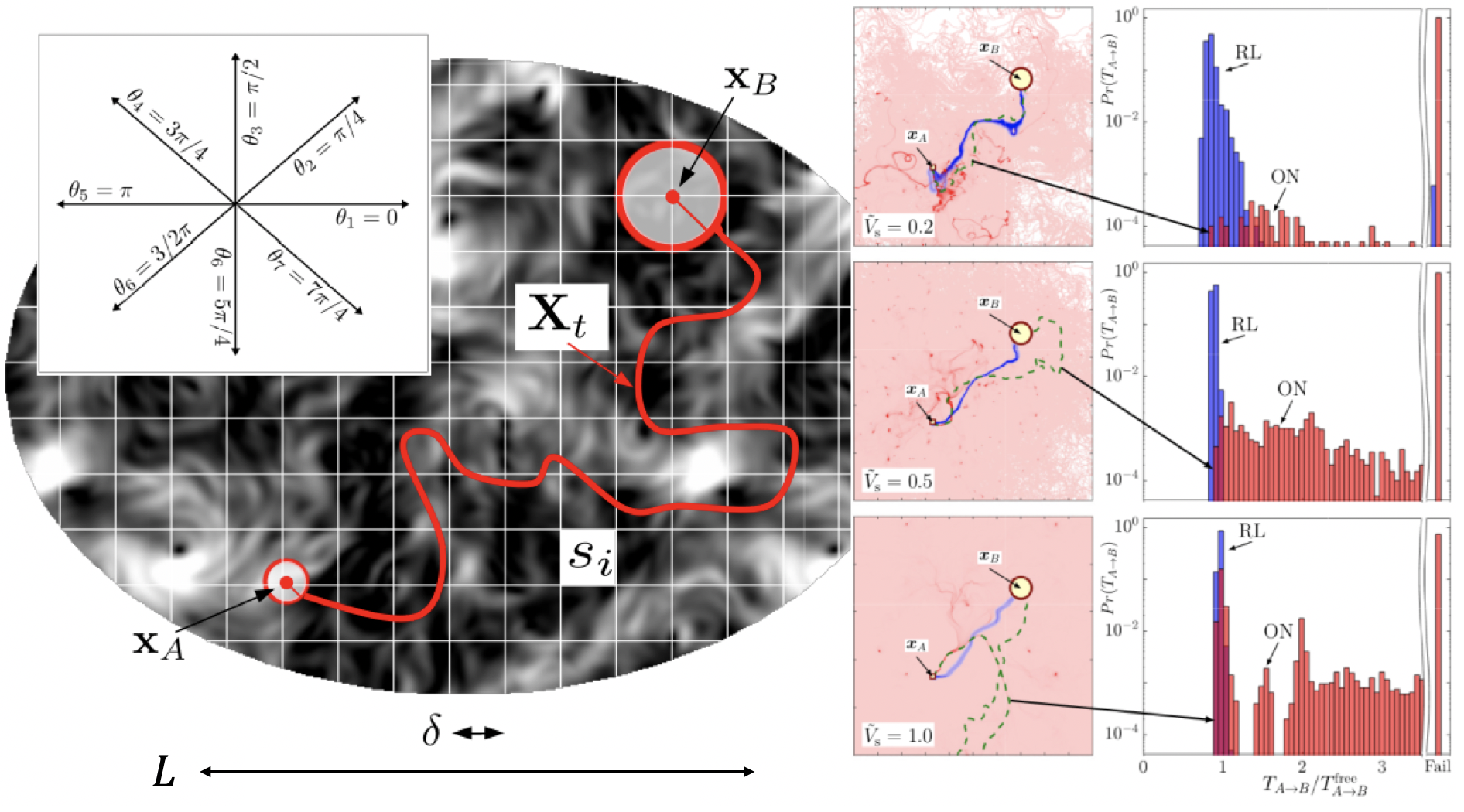}
\caption{Left: Image of one turbulent snapshot used as the advecting flow, with the starting,  ${\ve x}_{A}$, and ending point,  ${\ve x}_{B}$, of our  problem. We also show an illustrative navigation trajectory $\ve X_t$.  The flow is obtained from a spatially periodic snapshot of a 2D turbulent configuration in the inverse energy cascade regime with a multi-scale power-law  Fourier spectrum, $E(k) = \sum_{k< {\ve k} <k+1} |{\ve u}({\ve k})|^2 \sim k^{-5/3}$. For RL optimization, the initial conditions are taken randomly inside a circle of radius $d_A$ centered around ${\ve x}_{A}$. Similarly, the final target is the circle of radius $d_{B}$ centered around ${\ve x}_{B}$. The flow area is covered by a grid-world with tiles $s_i$ with $i=1,\dots,N_s$ and $N_s=900$ of size $\delta\times\delta$ which identify the state-space for the RL protocol. The large-scale periodicity of the underlying flow is $L$, and we fixed $\delta=L/10$. Every time interval $\Delta t$, the  unmanned  vessel selects one of the 8 possible actions $a_j$ with $j=1\dots 8$ (the steering directions $\theta_j$ depicted in left top inset)  according to a  policy $\pi(a|s)$, where $\pi$ is the probability distribution of the action $a$ given the current state $s$ of the agent at that time. \MB{The policy is optimized during the learning to maximize the total reward, $r_{tot}$, proportional to minus the navigation time, $r_{tot}\sim -T_{{\ve x}_{A} \rightarrow {\ve x}_{B}}$, so that the maximal reward corresponds to the fastest trajectory. For the policy to converge, the actor-critic method requires to accumulate experience over a number of the order of $1000$ different trajectories, with small variations depending on the values of the slip velocity $\tilde V_{\rm s}$ and the specific flow properties.} \obs{In a second series of experiments we added an additional action, the possibility to switch off the power, i.e. to let $V_s=0$. This allows the vessel to fully take advantage of the flow and save energy (see section \ref{sec:minimal}).} Right: spatial concentrations of trajectories for three values of $\tilde V_{\rm s}$. The flow region is color coded proportionally to the time the trajectories spend in each pixel area for both  ON  (red) and  RL (blue). Light colors refer to low occupation and bright to high occupation. The green-dashed line shows the best ON out the $20000$ trajectories. Right histograms:  arrival time  distribution for ON (red) and RL (blue). Probability of not reaching the target within the upper time limit is plotted in the {\it Fail} bar.}
\label{fig:initial}
\end{figure}
In this paper, we extend some of the results from a recent theoretical and numerical study \cite{biferale2019zermelo}, tackling Zermelo's problem for navigation in a two-dimensional fully
turbulent flow in the presence of an inverse energy cascade, i.e. with chaotic, multi-scale and rough velocity distributions \cite{alexakis2018cascades}, see Fig.~\ref{fig:initial} for a summary of the problem.  In such a flow, even for time-independent  configurations, trivial or naive navigation policies  can be extremely inefficient and ineffective if the set of actions by the vessel are limited. 
We show that an approach based on semi-supervised AI algorithms  using actor-critic  Reinforcement
Learning (RL) \cite{sutton2018} is able to find robust quasi-optimal stochastic policies that accomplish the task.  Furthermore, we compare RL with solutions from Optimal Navigation (ON) theory \cite{pontryagin2018mathematical} and show that the latter is of almost no practical use for the case of navigation in turbulent flows due to strong sensitivity to the initial (and final) conditions, in contrast to what happens for simpler  advecting flows \cite{schneider2019optimal}.
RL has shown to have promising potential to similar problems, such as the training of smart inertial particles or swimming objects navigating intense vortex regions~\cite{colabrese2018smart,colabrese2017flow,gustavsson2017finding}. 

We present here results from navigating one static snapshot of 2D turbulence (for time-dependent flows see \cite{biferale2019zermelo}).  In
Fig.~\ref{fig:initial} we show a sketch of the setup. Our goal is to find trajectories (if they
exist) that join the region close to $\bx_{A}$ with a
target close to $\bx_{B}$ in the shortest time, supposing that the vessels obey
the following equations of motion:
\begin{equation}
\begin{cases}
  \dot {\ve X}_t  = {{\ve u}({\ve X}_t,t)} + {\ve U}^{ctrl}({\ve X}_t) \\
  {\ve U}^{ctrl}({\ve X}_t) = V_{\rm s} {\bn}({\ve X}_t)
\label{eq:theta}
\end{cases}
\end{equation}
where {${\ve u}({\ve X}_t,t)$} is the velocity of the underlying 2D
advecting flow, and ${\ve U}^{ctrl}({\ve X}_t) = V_{\rm s}{\bn}({\ve X}_t)$ is the control slip
velocity of the vessel with fixed intensity $V_{\rm s}$ and varying
steering direction:
 ${\bn} ({\ve X}_t) = ( \cos[\theta_t], \sin[\theta_t] )$,
where the angle is evaluated along the trajectory, $\theta_t = \theta({\ve X}_t)$.
We introduce a dimensionless slip velocity by normalizing with the maximum velocity $u_{\max}$ of the underlying flow: $\tilde{V}_{\rm s} = V_{\rm s}/u_{\max}.$ Zermelo's problem reduces to optimize the steering direction  $\theta$ in order to reach the target~\cite{zermelo1931}. For time independent flows, optimal navigation (ON)  control theory gives a general solution\cite{techy2011optimal,mannarini2016visir}. Assuming that
the angle $\theta$ is controlled continuously in time, the optimal steering
angle must satisfy the following time-evolution:
\begin{equation}
\dot \theta_t = A_{21}\sin^2\theta_t - A_{12}\cos^2\theta_t + (A_{11} -
A_{22})\cos\theta_t \sin\theta_t\,, \label{eq:ON}
\end{equation}
where $A_{ij}=\partial_ju_i(\ve X_t)$ is evaluated along the agent
trajectory $\ve X_t$ obtained from Eq.~(\ref{eq:theta}). The set of equations (\ref{eq:theta}-\ref{eq:ON}) may lead to chaotic dynamics even for time-independent flows in two spatial dimensions. Due to the sensitivity to small perturbations in chaotic systems the ON approach becomes useless for many practical applications.

\section{Methods}
\label{sec:RL_methods}
RL applications \cite{sutton2018} are based on the idea that an optimal solution can be obtained by learning from continuous interactions of an agent with its environment. The agent interacts with the environment by sampling its states $s$, performing actions $a$ and collecting rewards $r$. In our case the vessel acts as the agent and the two-dimensional flow as the environment.
In the approach used here, actions are chosen randomly with a probability that is given by the policy
$\pi(a|s)$, given the current flow-state $s$. The goal is to find the optimal policy $\pi^*(a|s)$ that maximizes the total reward,
$ r_{\rm tot} = \sum_t r_t\,,$
accumulated along one episode. \MBB{For the purpose to find the fastest trajectory we used $r_t$ composed of three different terms;
\begin{equation}
    r_t = -\Delta t + \frac{|{\ve x}_B - {\ve X}_{t-\Delta t}|}{V_s} - \frac{|{\ve x}_B - {\ve X}_{t}|}{V_s}\,.
    \label{eq:reward-min-time}
\end{equation}
The first term accumulates a large penalty if it takes long for the agent to reach the end point, while the second and third terms \obs{describe the change in free-flight time to the target, i.e. the difference in time it would take, if the flow is neglected, to reach the target from the locations at this and the previous state change.}~\cite{art1}. It follows the the total reward is  proportional to minus the actual time taken by the trajectory to reach the target, $$r_{tot} \sim -T_{{\ve x}_{A} \rightarrow {\ve x}_{B}}\,,$$ 
neglecting a constant term that does not depend on the training, see~\cite{biferale2019zermelo} and Fig.~\ref{fig:initial} for precise definition of flow-states and agent-actions.}
An episode is finalized when the trajectory reaches the circle of radius $d_B$ around the target.
In order to converge to robust policies each episode is started with a uniformly random position within a given radius,  $d_A$,  from the starting point.
To estimate the expected total future reward  we follow  the one-step actor-critic method \cite{sutton2018} based on a gradient ascent in the policy parametrization.
\MBB{In the second part of our work, we modify the navigation setup by allowing the unmanned vessel to turn off its `engine', to allow it to navigate just following the flow without its own propulsion speed. \obs{In this framework, navigation can be optimal with respect to minimal energy consumption rather than time, or to a tradeoff between energy consumption and time.}
To repeat the training of the optimal policy taking into account of both aspects, energy and time, we modified our RL scheme as follows. First, we added the new action to turn off the vessel propulsion speed, i.e. letting $V_s=0$, in addition to the eight possible navigation angles considered before. Second, we modified the reward function in order to weigh \obs{the relative importance of navigation time and energy consumption. This was obtained by adding a new term describing the time the vessel consumes energy, $-\lambda \Delta t_{pow}$,} to the instantaneous reward in Eq.\eqref{eq:reward-min-time} as follows
\begin{equation}
    r_t = - (\Delta t + \lambda \Delta t_{pow}) + \frac{|{\ve x}_B - {\ve X}_{t-\Delta t}|}{V_s} - \frac{|{\ve x}_B - {\ve X}_{t}|}{V_s}\,.
    \label{eq:inst_reward_new}
\end{equation}
The total reward becomes proportional to minus the sum of the two time contributions,
\begin{equation}
r_{tot} \sim - \left (T_{{\ve x}_{A} \rightarrow {\ve x}_{B}} + \lambda T_{pow} \right )\,.
\label{eq:tot_rew_time_ene}
\end{equation}
\obs{The time $\Delta t_{pow}$ counts the time the vessel navigates with self propulsion, giving a total time $T_{pow}$ where energy is spent.
The factor $\lambda$ weighs the importance of energy consumption time and total navigation time in the optimisation.} 
We have repeated the training of the RL optimal policy with the new time-energy combined goals in the time-independent flow shown in Fig.~\ref{fig:initial}, as well as in a more realistic time-dependent 2D turbulent flow. The latter was obtained by solving the incompressible Navier-Stokes equations on a periodic square domain with side length $L=2\pi$ and $N=512^2$ number of collocation points, see~\cite{biferale2019zermelo} for more details about the flow.}

\section{Results (time-independent flows)}
\subsection{\obs{Shortest time, no energy constraints}}
In the right part of Fig. \ref{fig:initial} we show the main results comparing RL and ON approaches \cite{biferale2019zermelo}. The minimum time taken by the best trajectory to reach the target is of the same order for the two methods. The most important  difference between RL and ON lies in their robustness as seen by plotting the spatial density of trajectories in the right part of  
Fig.~\ref{fig:initial} for the optimal policies of ON and RL with three values of $\tilde V_{\rm s}$.
We observe that the RL trajectories (blue coloured area) form a much more coherent cloud in space, while the ON trajectories (red coloured area) fill space almost uniformly. Moreover, for small navigation velocities, many trajectories in the ON system approach regular attractors, as visible by the high-concentration regions. The rightmost histograms in  Fig.~\ref{fig:initial} show a comparison between the {probability} of arrival times for the trajectories illustrated in the two-dimensional domain, providing  a quantitative estimation
of the better robustness of RL compared to ON.
\MB{Other RL algorithms, such as Q-learning\cite{sutton2018}, could also be implemented and compared with other path search algorithms such as $A^*$ which is often used in computer science~\cite{russell2002artificial,lerner2009algorithmics}.}\\
\MBB{\subsection{Minimal energy consumption}
\label{sec:minimal}
In this section we present results on the simultaneous optimisation of minimal travel time and energy consumption. To begin with, we consider the same time-independent flow as in the previous section. 
\begin{figure}
\centering
\includegraphics[scale=0.42]{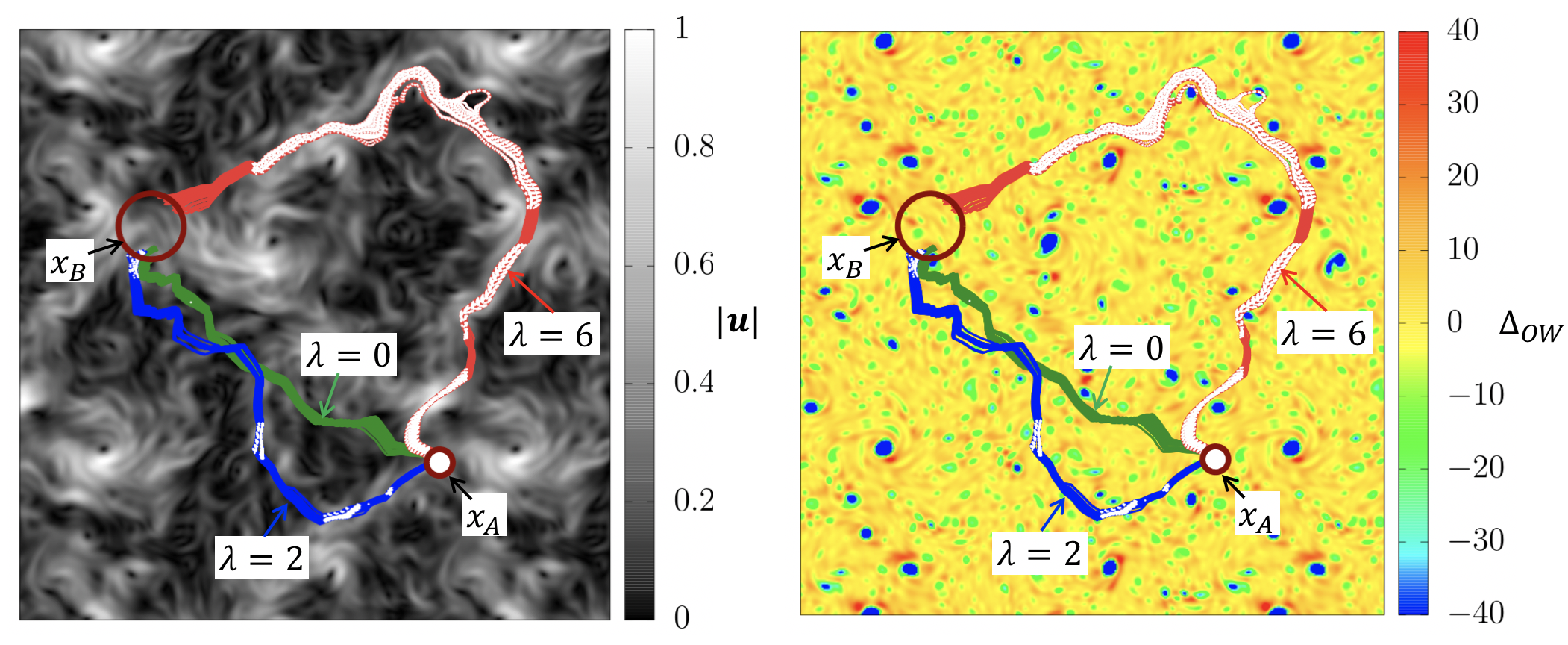}
\caption{Three sets of fifty trajectories going from point ${\ve x}_A$ to point ${\ve x}_B$, following the optimal policies for three values of $\lambda$ with propulsion speed \obs{either turned on, $\tilde{V_s} =0.8$ (color), or turned off, $V_s=0$ (white)}. (Left panel) The trajectories are plotted on top of the amplitude of the time-independent flow velocity, $|{\ve u}|$. (Right panel) Same trajectories plotted over the Okubo-Weiss parameter, $\Delta_{OW}$, see Eq.~(\ref{eq:Okubo_Weiss}).}
\label{fig:traj-OW}
\end{figure}
In Fig.~\ref{fig:traj-OW} we show three sets of trajectories following three policies obtained by optimising the reward (\ref{eq:tot_rew_time_ene}) for $\lambda=0$, $2$ and $6$.
The trajectories are superposed on the flow velocity amplitude $|{\ve u}({\ve X}_t,t)|$ (left panel) and the Okubo-Weiss parameter $\Delta_{OW}$~\cite{okubo1970horizontal,weiss1991dynamics} (right panel), defined as;
\begin{equation}
\Delta_{OW} = (A_{11}-A_{22})^2+(A_{21}+A_{12})^2-(A_{21}-A_{12})^2\,.
\label{eq:Okubo_Weiss}
\end{equation}
Here $A_{ij}$ is the fluid-gradient matrix as defined after Eq. (\ref{eq:ON}). The decomposition in Eq. (\ref{eq:Okubo_Weiss}) is particularly useful to distinguish strain dominated ($\Delta_{OW}>0$, orange-red colors) from vortex dominated ($\Delta_{OW}<0$, green-blue colors) regions of the flow. 
Colored regions of the trajectories show where the action is to have the propulsion on and white regions show where the propulsion is off.
When $\lambda=0$, the energy consumption does not matter for the reward, and the only difference compared to the case in the previous section is that the policy can now choose one additional action: the zero self propulsion speed. However, as seen from Fig.~\ref{fig:traj-OW}, this action is rarely chosen when $\lambda=0$, and the vessel navigates with a constant self-propelling velocity.
On the other hand, when the energy-dependent reward is activated, as in the case of $\lambda=2$, we observe a difference in the optimal path followed by the vessel. This is because it has to balance the penalties from the total navigation time and the time with self-propulsion. When $\lambda$ becomes larger, this difference in the optimal path becomes more significant. For $\lambda=6$ we observe trajectories that are much longer and dominated by passive navigation, just following the flow.
\begin{figure}
\centering
\includegraphics[scale=0.45]{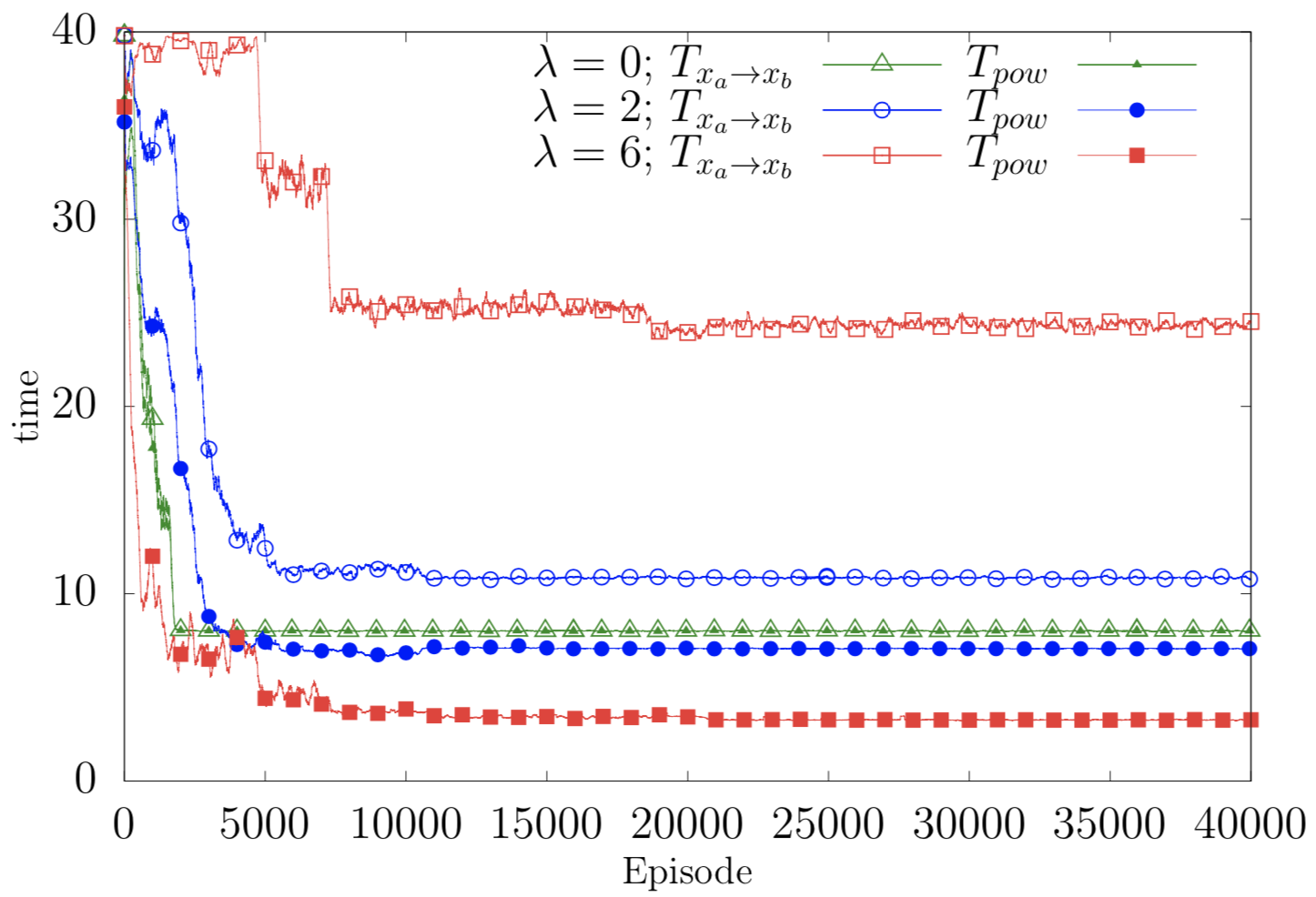}
\caption{Total navigation time, $T_{{\ve x}_{A} \rightarrow {\ve x}_{B}}$ (open symbols) and total power-on time, $T_{pow}$ (full symbols), measured for different trajectories obtained during the training as a function of the episode number, and for three different values of $\lambda$.}
\label{fig:rewards}
\end{figure}
To have a more accurate comparison of the arrival time to the target, $T_{{\ve x}_{A} \rightarrow {\ve x}_{B}}$, and the total active navigation time, $T_{pow}$, for the different values of $\lambda$, we show in Fig.~\ref{fig:rewards} the evolution of these two terms as functions of the episode number during the training of the three different policies. The total reward~\eqref{eq:tot_rew_time_ene} is a linear combination of these two terms, where $T_{pow}$ is multiplied by the factor $\lambda$. We first observe that the training converges after around 10k episodes. Second, we see that for $\lambda=0$, both $T_{{\ve x}_{A} \rightarrow {\ve x}_{B}}$ and $T_{pow}$ lies close to each other for all episodes, suggesting that the optimal policy never found a state where it is better to navigate with zero propulsion to reach the target faster. For values of $\lambda$ larger than zero, the found policies end up with $T_{pow}$ below the value of the $\lambda=0$ case,  with the consequence of saving energy even though the time to reach the target is longer. 
A final result for this case of time-independent flow is shown in Fig.~\ref{fig:pdf-time}, where we present the Probability Density Functions (PDFs) of the total navigation time, $T_{{\ve x}_{A} \rightarrow {\ve x}_{B}}$ (main panel) and of the power-on navigation time $T_{pow}$ (inset). The distributions are sampled over 40k trajectories with initial conditions close to ${\ve x}_{A}$ that follows the optimal policies obtained for five values of $\lambda$.
\begin{figure}
\centering
\includegraphics[scale=0.40]{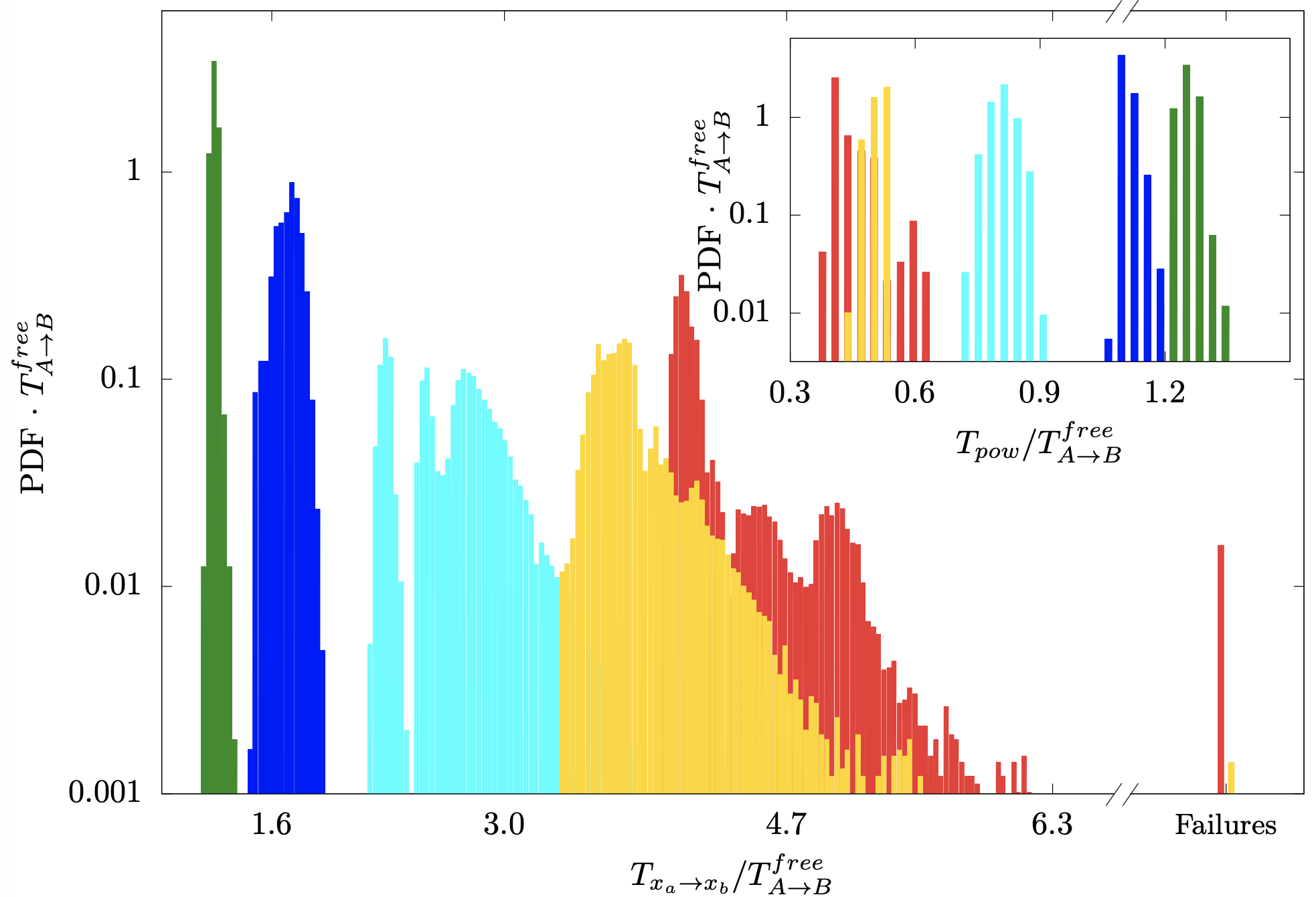}
\caption{(Main panel) PDFs of the arrival time, $T_{{\ve x}_{A} \rightarrow {\ve x}_{B}}$ normalized by the free-flight time $T_{A \rightarrow B}^{free}$, and measured over 40k different trajectories evolving on a time-independent flow. The different colors indicate different values of $\lambda$, from $\lambda=0$ (no extra cost for using power-on, green color) up to $\lambda=10$ (yellow color). The failures bars indicate the probability that a trajectory following a given policy does not reach the final target. (Inset) PDFs of the power-on time, $T_{pow}$, normalized by $T_{A \rightarrow B}^{free}$, measured along the same 40k trajectories shown in the main panel.}
\label{fig:pdf-time}
\end{figure}
These PDFs show that for $\lambda=0$, both times are of the order of $1.2 \, T_{A \rightarrow B}^{free}$, where $T_{A \rightarrow B}^{free} = |{\ve x}_B-{\ve x}_A|/V_s \sim 6.4$ is the free-flight time to go from point $A$ to point $B$ with a fixed self propulsion speed $V_s$ and without flow. For larger $\lambda$, the total navigation time increases while the power-on time decreases monotonically up to $\lambda=6$. Increasing $\lambda$ up to $10$ we do not observe further reduction of $T_{pow}$, the PDF only becomes more peaked around the value $\approx 0.4 \, T_{A \rightarrow B}^{free}$ as found for $\lambda=6$. This result suggests that we have found the minimal amount of propulsion required for the vessel to be able to navigate to the target. 
\section{Results (time-dependent flow)}
In this last section we consider the same optimal navigation problem as in the previous section, but with a more realistic time-dependent flow. For this case we adopted a small self-propulsion velocity, $\tilde{V}_s=0.2$, i.e. only $20\%$ of the maximal flow velocity amplitude. 
In Fig.~\ref{fig:time-dep-pdfs} we present, as in the previous section, the PDFs of both $T_{{\ve x}_{A} \rightarrow {\ve x}_{B}}$ (solid lines full symbols) and $T_{pow}$ (dashed lines empty symbols) obtained over 60k different trajectories following the converged optimal policies for $\lambda=0$, and $\lambda=2$. These results show that, as for the time-independent case, when $\lambda>0$ RL finds a solution that spends less energy at the cost of a longer total navigation time compared to the solution when $\lambda=0$.
Let us stress that with a probability of the order of $1$ in $1000$ we observed trajectories that were not able to reach the final target, as indicated by the failure bars reported in Fig.~\ref{fig:time-dep-pdfs}.
\begin{figure}
\centering
\includegraphics[scale=0.40]{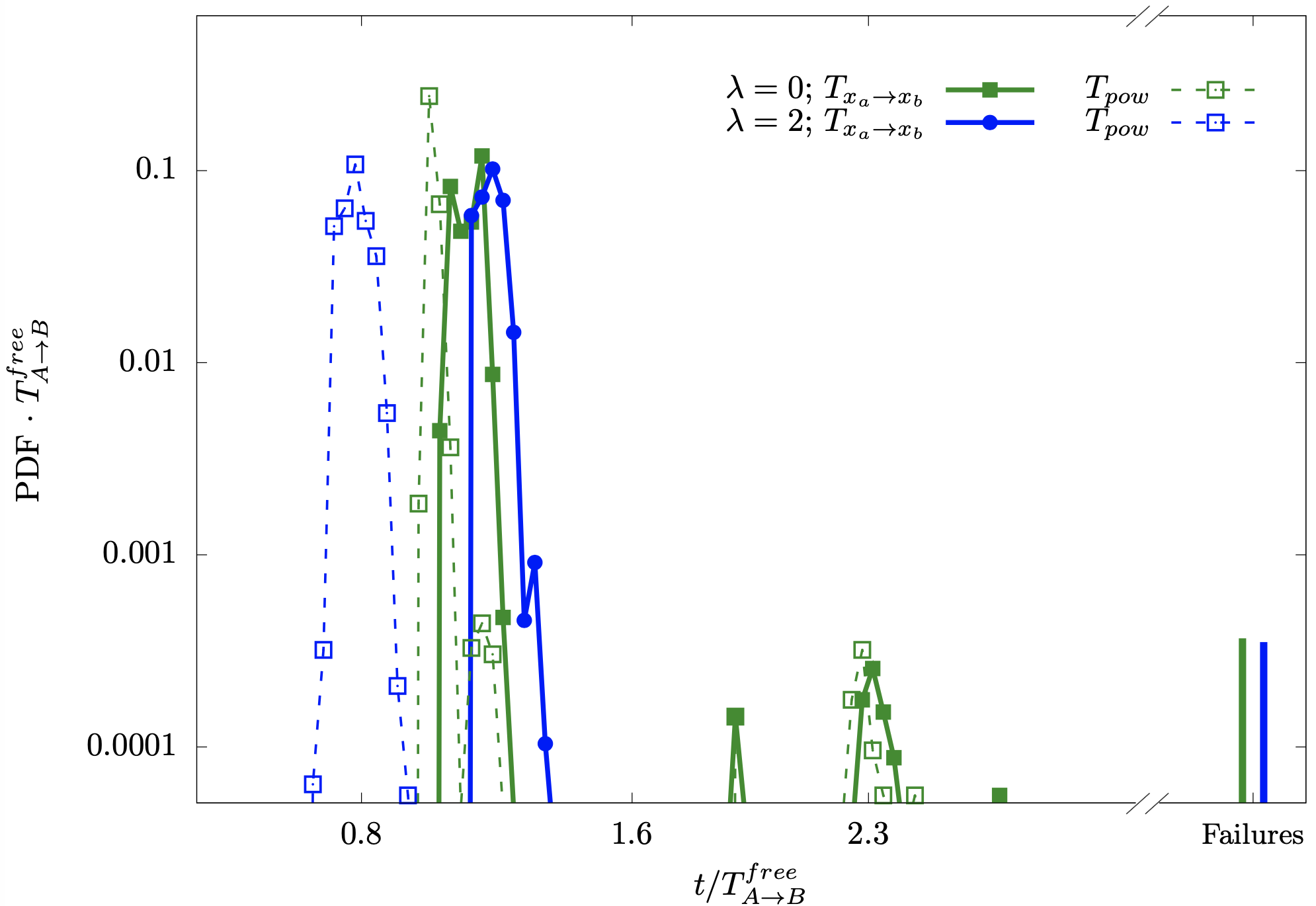}
\caption{PDFs of the arrival time, $T_{{\ve x}_{A} \rightarrow {\ve x}_{B}}$ (solid lines, full symbols) and of the power-on time, $T_{pow}$ (dashed lines, open symbols) measured over 60k different trajectories following the optimal policy in a time-dependent flow, both normalized by the free-flight time $T_{A \rightarrow B}^{free}$. Colours distinguish the two values of $\lambda$ used during the training. The navigation speed used along these trajectories is $\tilde{V}_s=0.2$, hence, $20\%$ of the maximum flow velocity amplitude. The failures bars indicate the probability of a trajectory to not reach the target after a long navigation time.}
\label{fig:time-dep-pdfs}
\end{figure}
Finally, Fig.~\ref{fig:time-dep-traj} shows six different snapshots at different times during the evolution of two different sets of trajectories that follows the optimal policies obtained for $\lambda=0$ and $\lambda=2$. The trajectories are superposed on the time-dependent flow velocity. Similar to Fig.~\ref{fig:traj-OW}, white regions on the trajectories show where the vessel is navigating with zero self propulsion speed. We remark that even when $\lambda=0$, the found optimal policy chooses the $V_s=0$ action in the region close to the target. As a result, the PDFs of the total navigation time and the power-on time are not identical even for the case of $\lambda=0$.
\obs{This is a very nice example of the fact that the resulting policy in RL benefits from the added control when the set of allowed actions is enlarged and that, in our particular application, passively moving with the flow can be better than navigating when the flow blows you in the right direction, independently of the requirement to minimize energy.}
\begin{figure}
\centering
\includegraphics[scale=0.45]{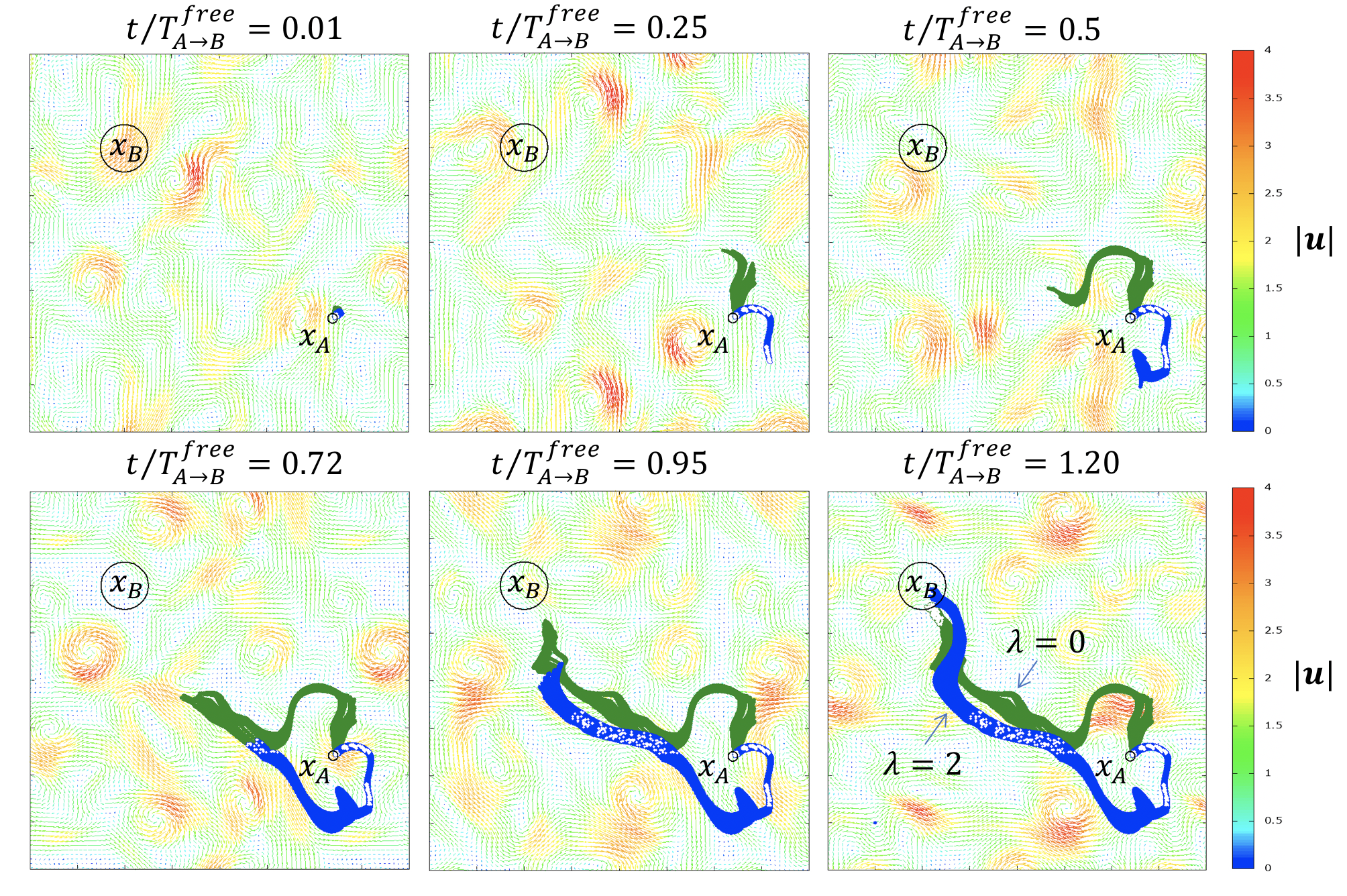}
\caption{Six snapshots at different times taken during the evolution of two sets of trajectories in a time-dependent flow, following the optimal policies for $\lambda=0$ (green color) and $\lambda=2$ (blue color). The six times are normalized to the free flight time. The flow streamlines are coloured proportionally to their amplitude, while the white points along the navigation trajectories indicate locations where the selected action was passive navigation, i.e. $V_S=0$.}
\label{fig:time-dep-traj}
\end{figure}}

\section{Conclusions}
We have \MBB{first} discussed a  systematic investigation of Zermelo's time-optimal navigation problem in a realistic 2D turbulent flow, comparing both  RL and ON approaches \cite{biferale2019zermelo}. We showed that RL stochastic algorithms are key to bypass unavoidable instability given by the chaoticity of the environment and/or by the strong sensitivity of ON on the initial conditions in the presence of non-linear flow configurations. RL methods offer also a wider flexibility, being applicable to energy-minimization problems and in situations where the flow evolution is known only in a statistical sense as in partially observable Markov processes.
\MB{Let us stress that, instead of starting from a completely random policy as we did here, it is also possible to implement RL to improve a-priori policies designed for a particular problem. For example, one can use an RL approach to optimize an initial {\it trivial policy}, where the navigation angle is selected as the action that points most directly toward the target.}
\MBB{In the second part of this work, we further analyzed the more complex problem where the optimization of the total navigation time \obs{is balanced by} the energy consumption required to reach the target. Also in this case, we found that RL is able to converge to non-trivial solutions where the vessel navigates most of the time as a passive object transported by the flow, with only a minimum number of corrections to its trajectory required to reach the final target. 
}

\begin{acknowledgments}
L.B. and M.B. acknowledge funding from the European Union Programme (FP7/2007-2013) AdG ERC grant  No.339032. K.G. acknowledges funding from the Knut and Alice Wallenberg Foundation, Grant No. KAW 2014.0048, and Vetenskapsrådet, Grant No. 2018-03974. F.B acknowledges funding from the European Research Council under the European Union’s Horizon 2020 Framework Programme (No. FP/2014-2020) ERC Grant Agreement No.739964 (COPMAT).
\end{acknowledgments}

\bibliography{bibliography}

\appendix

\end{document}